\documentclass[letterpaper,twocolumn,10pt]{article}
\usepackage{usenix2019_v3,epsfig,endnotes}
\usepackage{amsmath}
\usepackage[linesnumbered,ruled]{algorithm2e}
\usepackage{booktabs}
\usepackage{graphicx}
\usepackage{listings} % to list code snippets
\usepackage[defaultmathsizes,italic]{mathastext}
\usepackage{multirow}
\usepackage[normalem]{ulem}
\usepackage{setspace}
\usepackage{siunitx}
\usepackage{times}
\usepackage{url}
\usepackage{xcolor}

\def\cfigure[#1,#2,#3]{
\begin{figure}
\vspace*{0mm}
\begin{center}

\includegraphics[width=\linewidth]{#1}

\vspace*{0mm}\caption[]{#2
} \label{#3}

\vspace*{-1mm}
\end{center}
\end{figure}}

\def\cfigureslim[#1,#2,#3]{
\begin{figure}
\vspace*{0mm}
\begin{center}

\includegraphics[width=.9\linewidth]{#1}

\vspace*{0mm}\caption[]{#2
} \label{#3}

\vspace*{-1mm}
\end{center}
\end{figure}}

\def\cfigurefour[#1,#2,#3]{
\begin{figure}
\vspace*{0mm}
\begin{center}

\includegraphics[width=4in]{#1}

\vspace*{-3mm}\caption[]{#2
} \label{#3}

\vspace*{-5mm}
\end{center}
\end{figure}}

\def\wfigure[#1,#2,#3]{
\begin{figure*}
\vspace*{0mm}
\begin{center}
 \includegraphics[width=\textwidth]{#1}
 \vspace*{-3mm}\caption[]{#2
} \label{#3}

\end{center}
\end{figure*}}

\def\wfigurex[#1,#2,#3]{
\begin{figure*}
\vspace*{0mm}
\begin{center}
 \includegraphics[width=0.9\textwidth]{#1}
 \vspace*{-3mm}\caption[]{#2
} \label{#3}
\end{center}
\end{figure*}}

\def\threefigure[#1,#2,#3,#4,#5]{
\begin{figure*}
\vspace*{0mm}
\begin{center}

\begin{tabular}{ccc}
\includegraphics[width=2in]{#1} & \includegraphics[width=2in]{#2} &  \includegraphics[width=2in]{#3} \\
(a) & (b) & (c) \\ \\
\end{tabular}

\vspace*{-3mm}\caption[]{#4
} \label{#5}

\vspace*{-5mm}
\end{center}
\vspace*{-2mm}
\end{figure*}}

\def\dcfigure[#1,#2,#3,#4,#5,#6]{
{
\begin{figure*}
\begin{center}
\begin{minipage}[c]{\columnwidth}{
\includegraphics[width=\columnwidth]{#1}
\vspace*{0mm}\caption[]{#2} \label{#3} \
}\end{minipage}\hspace*{\columnsep}\
\begin{minipage}[c]{\columnwidth}{
\includegraphics[width=\columnwidth]{#4}
\vspace*{0mm}\caption[]{#5}\label{#6} \
}\end{minipage}
\end{center}
\end{figure*}
}
}

\def\tableByTable[#1,#2,#3,#4,#5,#6]{
{
\begin{table*}
\begin{center}
\begin{minipage}[c]{3in}{
\centering
{#1}
\vspace*{0mm}\tabcaption[]{#2}\label{#3} \
}\end{minipage}\hspace*{\columnsep}\
\begin{minipage}[c]{3in}{
\centering
{#4}
\vspace*{0mm}\tabcaption[]{#5}\label{#6} \
}\end{minipage}
\end{center}
\end{table*}
}
}

\def\figureByTable[#1,#2,#3,#4,#5,#6]{
{
\begin{figure*}
\begin{center}
\begin{minipage}[c]{3in}{
\centering
\includegraphics[width=\textwidth]{#1}
\vspace*{0mm}\figcaption[]{#2} \label{#3} \
}\end{minipage}\hspace*{\columnsep}\
\begin{minipage}[c]{3.3in}{
\centering
{#4}
\vspace*{0mm}\tabcaption[]{#5}\label{#6} \
}\end{minipage}
\end{center}
\end{figure*}
}
}

\def\tableByFigure[#1,#2,#3,#4]{
{
\begin{figure*}
\begin{center}
\begin{minipage}[c]{4.0in}{
\centering
\small
{#1}
}\end{minipage}\hspace*{\columnsep}\
\begin{minipage}[c]{3.1in}{
\centering
\includegraphics[width=2.6in]{#2}
}\end{minipage}
\end{center}
\vspace*{-0.0in}\caption[]{#3}\label{#4}
\end{figure*}
}
}

\def\doublecfigure[#1,#2,#3,#4]{
{
\begin{figure}
\begin{center}
\begin{minipage}[c]{1.5in}{
\begin{center}
\includegraphics[width=1.5in]{#1}%\\(a)
\end{center}
}\end{minipage}\hspace*{1em}\
\begin{minipage}[c]{1.5in}{
\begin{center}
\includegraphics[width=1.5in]{#2}%\\(b)
\end{center}
}\end{minipage}
\vspace*{0mm}\caption[]{#3} \label{#4} \
\end{center}
\end{figure}
}
}

\def\qcfigure[#1,#2,#3,#4,#5,#6]{
{
\begin{figure*}
\vspace*{0.2in}\
\begin{center}
\begin{minipage}[c]{3in}{
\includegraphics[width=3in]{#1}
\vspace*{-3mm}
}
\end{minipage}\hspace*{0.5in}\
\begin{minipage}[c]{3in}{
\includegraphics[width=3in]{#2}
\vspace*{-3mm}
}\end{minipage}

\begin{minipage}[c]{3in}{
\includegraphics[width=3in]{#3}
\vspace*{-3mm}
}
\end{minipage}\hspace*{0.5in}\
\begin{minipage}[c]{3in}{
\includegraphics[width=3in]{#4}
\vspace*{-3mm}
}\end{minipage}
\end{center}
\caption[]{#5}\label{#6}
\end{figure*}
}
}

\def\twfigure[#1,#2,#3,#4,#5]{
{
\begin{figure*}
\vspace*{0.2in}\
\begin{center}
\begin{minipage}[c]{6.5in}{
\includegraphics[width=6.5in]{#1}
\vspace*{-3mm}
}
\end{minipage}

\begin{minipage}[c]{6.5in}{
\includegraphics[width=6.5in]{#2}
\vspace*{-3mm}
}\end{minipage}

\begin{minipage}[c]{6.5in}{
\includegraphics[width=6.5in]{#3}
\vspace*{-3mm}
}
\end{minipage}
\end{center}
\caption[]{#4}\label{#5}
\end{figure*}
}
}

\def\dwfigure[#1,#2,#3,#4]{
{
\begin{figure*}
\vspace*{0.2in}\
\begin{center}
\begin{minipage}[c]{6.5in}{
\includegraphics[width=6.5in]{#1}
\vspace*{-3mm}
}
\end{minipage}

\begin{minipage}[c]{6.5in}{
\includegraphics[width=6.5in]{#2}
\vspace*{-3mm}
}\end{minipage}

\end{center}
\caption[]{#3}\label{#4}
\end{figure*}
}
}

\def\dssfigure[#1,#2,#3,#4,#5,#6]{
{
\begin{figure*}
\vspace*{0.2in}\
\begin{center}
\begin{minipage}[c]{4in}{
\includegraphics[width=4in]{#1}
\vspace*{-3mm}\caption[]{#2} \label{#3} \
}\end{minipage}\hspace*{0.5in}\
\begin{minipage}[c]{2in}{
\includegraphics[width=2in]{#4}
\vspace*{-3mm}\caption[]{#5}\label{#6} \
}\end{minipage}
\end{center}
\vspace*{-0.4in}\
\end{figure*}
}
}

\def\dsfigure[#1,#2,#3,#4,#5,#6]{
{
\begin{figure*}
\vspace*{0.2in}\
\begin{center}
\begin{minipage}[c]{3in}{
\includegraphics[width=3in]{#1}
\vspace*{-3mm}\caption[]{#2} \label{#3} \
}\end{minipage}\hspace*{0.5in}\
\begin{minipage}[c]{3in}{
\hspace*{0.5in}\
\includegraphics[height=3in]{#4}
\vspace*{-3mm}\caption[]{#5}\label{#6} \
}\end{minipage}
\end{center}
\vspace*{-0.4in}\
\end{figure*}
}
}

\def\dsyfigure[#1,#2,#3,#4,#5,#6]{
{
\begin{figure*}
\vspace*{0.2in}\
\begin{center}
\begin{minipage}[c]{2.5in}{
\includegraphics[height=2.5in]{#1}
\vspace*{-3mm}\caption[]{#2} \label{#3} \
}\end{minipage}\hspace*{0.5in}\
\begin{minipage}[c]{2.5in}{
\includegraphics[height=2.5in]{#4}
\vspace*{-3mm}\caption[]{#5}\label{#6} \
}\end{minipage}
\end{center}
\vspace*{-0.4in}\
\end{figure*}
}
}

\def\dyfigure[#1,#2,#3,#4,#5,#6]{
{
\begin{figure*}
\vspace*{0.2in}\
\begin{center}
\begin{minipage}[c]{3in}{
\includegraphics[height=3in]{#1}
\vspace*{-3mm}\caption[]{#2} \label{#3} \
}\end{minipage}\hspace*{0.5in}\
\begin{minipage}[c]{3in}{
\includegraphics[height=3in]{#4}
\vspace*{-3mm}\caption[]{#5}\label{#6} \
}\end{minipage}
\end{center}
\vspace*{-0.4in}\
\end{figure*}
}
}

\def\dyoldfigure[#1,#2,#3,#4,#5,#6]{
{
\begin{figure*}
\vspace*{0.2in}\
\begin{center}
\begin{minipage}[c]{3in}{
\epsfysize=2.0in\
\hspace{0.5in}\
\epsfbox{#1}
\vspace*{-3mm}\caption[]{#2} \label{#3} \
}\end{minipage}\hspace*{0.25in}\
\begin{minipage}[c]{3in}{
\epsfysize=2.0in\
\hspace{0.5in}\
\epsfbox{#4}
\vspace*{-3mm}\caption[]{#5}\label{#6} \
}\end{minipage}
\end{center}
\vspace*{-0.4in}\
\end{figure*}
}
}

\def\cfiguredouble[#1,#2,#3,#4]{
\begin{figure}
\vspace*{0.2in}\
\begin{center}
\begin{minipage}[c]{1.5in}{
\epsfxsize=1.5in\
\epsfbox{#1}
}\end{minipage}\hspace*{0.1in}\
\begin{minipage}[c]{1.5in}{
\epsfxsize=1.5in\
\vspace{0.1in}\epsfbox{#2}
}\end{minipage}\vspace*{-0.10in} \caption[]{#3}\label{#4}
\end{center}
\vspace*{-0.4in}\
\end{figure}
}

\def\wpfigure[#1,#2,#3,#4]{
\begin{figure*}
\vspace*{4mm}
\begin{center}

\includegraphics[width=#4]{#1}

\vspace*{-3mm}\caption[]{#2
} \label{#3}

\vspace*{-5mm}
\end{center}
\end{figure*}}

\def\wprfigure[#1,#2,#3,#4,#5]{
\begin{figure*}
\vspace*{4mm}
\begin{center}

\includegraphics[width=#4, angle=#5]{#1}

\vspace*{-3mm}\caption[]{#2
} \label{#3}

\vspace*{-5mm}
\end{center}
\end{figure*}}

\def\DoubleFigureWSlide[#1,#2,#3,#4,#5,#6,#7,#8,#9]{
\begin{figure*}
\vspace*{#9}
\begin{center}
\begin{minipage}{#4}
\includegraphics[width=#4]{#1}
\vspace*{-3mm}\caption{#2
}\label{#3}
\end{minipage}
\hspace{2em}
\begin{minipage}{#8}
\includegraphics[width=#8]{#5}
\vspace*{-3mm}\caption{#6
}\label{#7}
\end{minipage}
\vspace*{-5mm}
\end{center}
\end{figure*}
}

\def\DoubleFigureW[#1,#2,#3,#4,#5,#6,#7,#8]{
\begin{figure*}
\vspace*{0in}
\begin{center}
\begin{minipage}{#4}
\includegraphics[width=#4]{#1}
\vspace*{-3mm}\caption{#2
}\label{#3}
\end{minipage}
\hspace{2em}
\begin{minipage}{#8}
\includegraphics[width=#8]{#5}
\vspace*{-3mm}\caption{#6
}\label{#7}
\end{minipage}
\vspace*{-5mm}
\end{center}
\end{figure*}
}

\def\DoubleFigureWHack[#1,#2,#3,#4,#5,#6,#7,#8]{
\begin{figure*}
\vspace*{0in}
\begin{center}
\begin{minipage}{3in}
\includegraphics[width=#4]{#1}
\vspace*{-3mm}\caption{#2
}\label{#3}
\end{minipage}
\hspace{2em}
\begin{minipage}{3in}
\includegraphics[width=#8]{#5}
\vspace*{-3mm}\caption{#6
}\label{#7}
\end{minipage}
\vspace*{-5mm}
\end{center}
\end{figure*}
}

\def\ddcfigure[#1,#2,#3,#4]{
\begin{figure*}
\vspace*{0.2in}\
\begin{center}
\begin{minipage}[c]{\columnwidth}{
\includegraphics[width=\columnwidth]{#1}
}\end{minipage}\hspace{0.5in}\
\begin{minipage}[c]{\columnwidth}{
\includegraphics[width=\columnwidth]{#2}
}\end{minipage} \caption[]{#3}\label{#4}
\end{center}
\end{figure*}
}

\def\ddcfigureSlide[#1,#2,#3,#4,#5]{
\begin{figure*}
\vspace*{#5}\
\begin{center}
\begin{minipage}[c]{3in}{
\includegraphics[height=3in]{#1}
}\end{minipage}\hspace{0.5in}\
\begin{minipage}[c]{3in}{
\includegraphics[height=3in]{#2}
}\end{minipage}\vspace*{-0.10in} \caption[]{#3}\label{#4}
\end{center}
\vspace*{-0.4in}\
\end{figure*}
}

\def\cxfigure[#1,#2,#3]{
\begin{figure}
\vspace*{4mm}
\begin{center}

\epsfxsize=2.5in\
\epsfbox{#1}\

\vspace*{-0.10in}\caption[]{#2
} \label{#3}

\vspace*{-5mm}
\end{center}
\vspace*{-2mm}
\end{figure}}

\newcommand{\boldparagraph}[1]{\vspace*{1ex}\noindent\textbf{#1}\hspace{1em}}

\newcommand{\etal}{\textit{et al.}}

\newcommand{\figtitle}[1]{#1 --}

\newcommand{\ignore}[1]{}

\newcommand{\us}[1]{$\mu$s}

\newcommand{\x}[1]{$\times$}

\newcommand{\reffig}[1]{Figure~\ref{#1}}
\newcommand{\reftab}[1]{Table~\ref{#1}}
\newcommand{\reflst}[1]{Listing~\ref{#1}}
\newcommand{\refsec}[1]{Section~\ref{#1}}

\newcommand{\myitem}[1]{\item \textbf{#1}}
\lstdefinestyle{cstyle}{
  float=tp,
  floatplacement=tbp,
  abovecaptionskip=5pt,
  breaklines=true,
  captionpos=b,
  frame=lines,
  xleftmargin=4pt,
  language=C,
  showstringspaces=false,
  basicstyle=\footnotesize\ttfamily,
  identifierstyle=\color{black},
  keywordstyle=\bfseries\color{green!40!black},
  commentstyle=\color{gray},
  stringstyle=\color{orange!70!black},
  morekeywords={size_t, uint64_t, PMEMoid, PMEMobjpool}
}

\newcommand{\DAXmmap}{DAX-\texttt{mmap()}}
\newcommand{\DAXmmapd}{DAX-mapped}

\newcommand{\FTLib}{Pangolin}

\newcommand{\lpo}{\texttt{libpmemobj}}
\newcommand{\Lpo}{\texttt{Libpmemobj}}

\newcommand{\Po}{Pmemobj}
\newcommand{\Por}{Pmemobj-R}

\newcommand{\mbuf}{micro-buffer}
\newcommand{\mbufs}{micro-buffers}
\newcommand{\Mbuf}{Micro-buffer}

\newcommand{\mbufed}{micro-buffered}

\newcommand{\mbufing}{micro-buffering}
\newcommand{\Mbufing}{Micro-buffering}

\newcommand{\pmemoid}{\texttt{PMEMoid}}

\newcommand{\clwb}{\texttt{CLWB}}
\newcommand{\sfence}{\texttt{SFENCE}}
\newcommand{\sigbus}{\texttt{SIGBUS}}

\begin{document}

\title{\bf \FTLib{}: A Fault-Tolerant Persistent Memory Programming Library}
\author{
  {\rm Lu Zhang}\\
  University of California, San Diego\\
  \textit{luzh@eng.ucsd.edu}
  \and
  {\rm Steven Swanson}\\
  University of California, San Diego\\
  \textit{swanson@cs.ucsd.edu}
} % end author

\maketitle

\begin{abstract}
Non-volatile main memory (NVMM) allows programmers to build complex, persistent,
pointer-based data structures that can offer substantial performance gains over
conventional approaches to managing persistent state. This programming model
removes the file system from the critical path which improves performance, but
it also places these data structures out of reach of file system-based fault
tolerance mechanisms (e.g., block-based checksums or erasure coding). Without
fault-tolerance, using NVMM to hold critical data will be much less attractive.

This paper presents \FTLib{}, a fault-tolerant persistent object library
designed for NVMM. \FTLib{} uses a combination of checksums, parity, and
micro-buffering to protect an application's objects from both media errors and
corruption due to software bugs. It provides these protections for objects of
any size and supports automatic, online detection of data corruption and
recovery. The required storage overhead is small (1\% for gigabyte-sized pools
of NVMM). \FTLib{} provides stronger protection, requires orders of magnitude
less storage overhead, and achieves comparable performance relative to the
current state-of-the-art fault-tolerant persistent object library.
\end{abstract}

\section{Introduction}
\label{sec:intro}

Emerging non-volatile memory (NVM) technologies (e.g., battery-backed NVDIMMs
\cite{micron-nvdimm} and 3D XPoint \cite{3dxpoint}) provide persistence with
performance comparable to DRAM. Non-volatile main memory (NVMM), is
byte-addressable, cache-coherent NVM that resides on the system's main memory bus.
The combination of NVMM and DRAM enables hybrid memory systems that offer
the promise of dramatic increases in storage performance and a more flexible
programming model.

A key feature of NVMM is support for direct access, or DAX, that lets
applications perform loads and stores directly to a file that resides in NVMM.
DAX offers the lowest-possible storage access latency and
enables programmers to craft complex, customized data structures for
specific applications. To support this model, researchers and industry have
proposed various persistent object systems
~\cite{nvheaps,mnemosyne,pmdk,dudetm,lognvmm,kaminotx,justdo,romulus}.

Building persistent data structures presents a host of challenges, particularly
in the area of crash consistency and fault tolerance. Systems that use NVMM must
preserve crash-consistency in the presence of volatile caches, out-of-order 
execution, software bugs, and system failures. To address these challenges, many 
groups have proposed crash-consistency solutions based on
hardware~\cite{wsp,thynvm,whisper,hwundoredo},
file systems~\cite{bpfs,pmfs,nova,novafortis},
user-space data structures and libraries~
\cite{nvtree,slottedpaging,nvheaps,mnemosyne,pmdk,lognvmm,romulus}, and
languages~\cite{pynvm,nvlc}.

Fault tolerance has received less attention but is equally important: To be
viable as an enterprise-ready storage medium, persistent data structures must include
protection from data corruption.  Intel processors report uncorrectable memory
media errors via a machine-check exception and the kernel forwards it
to user-space as a \sigbus{} signal. To our knowledge,
Xu~\etal{}~\cite{novafortis} were the first to design an NVMM file system that
detects and attempts to recover from these errors.  Among programming libraries,
only \lpo{} provides any support for fault tolerance, but it incurs 100\% space
overhead, only protects against media errors (not software ``scribbles''), and
cannot recover corrupted data without taking the object store offline.

Xu~\etal{} also highlighted a fundamental conflict between
\DAXmmap{} and file system-based fault tolerance: By design, \DAXmmap{} leaves
the file system unaware of updates made to the file, making it impossible for
the file system to update the redundancy data for the file.
Their solution is to disable file data protection while the file is mapped
and restore it afterward. This provides well-defined protection guarantees but
leaves file data unprotected when it is in use.

Moving fault-tolerance to user-space NVMM libraries solves this problem, but
presents challenges since it requires integrating fault tolerance into
persistent object libraries that manage potentially millions of small,
heterogeneous objects.

To satisfy the competing requirements placed on NVMM-based, \DAXmmapd{} object
store, a fault-tolerant persistent object library should provide at least the
following characteristics:

\begin{enumerate}

  \myitem{Crash-consistency}. The library should provide the means to ensure
  consistency in the face of both system failures and data corruption.
  
  \myitem{Protection against media and software errors}. Both types of errors
  are real threats to data stored to NVMM, so the library should provide
  protection against both.

  \myitem{Low storage overhead}. NVMM is expensive, so minimizing storage
  overhead of fault tolerance is important.
  
  \myitem{Online recovery}. For good availability, detection and recovery must
  proceed without taking the persistent object store offline.
  
  \myitem{High performance}. Speed is a key benefit of NVMM. If
  fault-tolerance incurs a large performance penalty, NVMM will be much less 
  attractive.

  \myitem{Support for diverse objects}. A persistent object system must support
  objects of size ranging from a few cache lines to many megabytes. 

\end{enumerate}

This paper describes \emph{\FTLib{}}, the first persistent object library to
satisfy all these criteria. \FTLib{} uses a combination of parity, replication,
and object-level checksums to provide space-efficient, high-performance fault
tolerance for complex NVMM data structures. \FTLib{} also introduces a new
technique for accessing NVMM called \emph{micro-buffering} that simplifies
transactions and protects NVMM data structures from programming errors.

We evaluate \FTLib{} using a suite of benchmarks and compare it to \lpo{}, a
persistent object library that offers a simple replication mode for fault
tolerance. Compared to \lpo{}, performance is similar, and \FTLib{} provides
stronger protection, online recovery, and greatly reduced storage overhead
(1\% instead of 100\%).

The rest of the paper is organized as follows:  \refsec{sec:background} provides a primer on
NVMM programming and NVMM error handling in Linux. \refsec{sec:design} describes
how \FTLib{} organizes data, manages transactions, and detects and repairs errors.
\refsec{sec:eval} presents our evaluations. \refsec{sec:related}
discusses related work. Finally, \refsec{sec:conclusion} concludes.

\section{Background} \label{sec:background}

\FTLib{} lets programmers build fault-tolerant, crash-consistent data structures
in NVMM. This section first introduces NVMM and the DAX mechanism applications
use to gain direct access to persistent data. Then, we describe the NVMM error
handling mechanisms that Intel processors and Linux provide.  Finally, we
provide a brief primer on NVMM programming using \lpo{}~\cite{pmdk}, the library on which \FTLib{} is based.

\subsection{Non-volatile Main Memory and DAX}

Several technologies are poised to make NVMM common in computer systems. 3D
XPoint~\cite{3dxpoint} is the closest to wide deployment. Phase change
memory (PCM), resistive RAM (ReRAM), and spin-torque transfer RAM (STT-RAM) are
also under active development by memory manufacturers. Flash-backed DRAM is
already available and in wide use. Linux and Windows both have support for
accessing NVMM and using it as storage media.

The performance and cost parameters of NVMM lie between DRAM and SSD. Its write
latency is longer than DRAM, but it will cost less per bit.  From the
storage perspective, NVMM is faster but more expensive than SSD.

The most efficient way to access NVMM is via direct access
(DAX)~\cite{intel2017intro} memory mapping (i.e., \DAXmmap{}).  To use
\DAXmmap{}, applications map pages of a file in an NVMM-aware file
system into their address space, so the application can access
persistent data from the user-space using load and store instructions, without
the file system intervening.

% \DAXmmap{} gives applications the fastest possible access to stored data,
% since it entirely eliminates file and operating system overheads. It also
% enables programmers to build complex data structures and store them directly
% in the mapped NVMM. These data structures persist across application restarts
% and system power failures.

\subsection{Handling NVMM Media Errors} \label{sec:errmodel}

To recover from data corruption, \FTLib{} relies on error detection and media
management facilities that the processor and operating system provide together. Below,
we describe these facilities available on Intel and Linux platforms. Windows
provides similar mechanisms.

\boldparagraph{Hardware Error Correction} Memory controllers for commercially
available NVMMs (i.e., battery-backed DRAM and 3D~XPoint) implement
error-correction code (ECC) in hardware to detect and correct media errors
when they can, and they report uncorrectable (but detectable) errors with a
machine check exception (MCE)~\cite{intel2017isa} that the operating system can
catch and attempt to handle.

\FTLib{} provides a layer of protection in addition to the ECC hardware
provides, but it does not require hardware ECC.
\FTLib{} uses checksums to detect errors that hardware cannot detect.
This mechanism also catches software bugs (which are
invisible to hardware ECC). ECC does,
however, improve performance by transparently handling many media errors.

Regardless of the ECC algorithm hardware provides, field studies of DRAM and
SSDs~\cite{dram-errors-wild, cosmic, dram-errors-modern, sector-errors,
flash-failures, ssd-failures} have shown that detectable but uncorrectable media
errors occur frequently enough to warrant additional software protection.
Furthermore, file systems~\cite{zhang2010end,novafortis,waflmetaprot} apply
checksums to their data structures to protect against scribbles.

\boldparagraph{Repairing Errors} When the hardware detects an uncorrectable
error, the Linux kernel marks the region surrounding the failed load as
``poisoned,''  and future loads from the region will fail with a bus error.
\FTLib{} assumes an error
poisons a 4~KB page since Linux currently manages memory failures at page
granularity.

If a running application causes an MCE (by loading from a poisoned page), the
kernel sends it a \sigbus{} and the application can extract the affected
address from the data structure describing the signal.

The software can repair the poisoned page by writing new data to the region. In
response, the operating system and NVDIMM firmware work together to remap the poisoned
addresses to functioning memory cells. The details of this process are part of
the Advanced Configuration and Power Interface (ACPI)~\cite{acpi62} for NVDIMMs.

Recent kernel patches~\cite{libnvdimm412,libnvdimm413,lkml-mce,pagepoison} and
NVMM library \cite{pmdk} provide utilities for user-space applications to
restore lost data by re-writing affected pages with recovered contents (if
available).

\subsection{NVMM Programming} \label{sec:pmemobj}

\cfigure[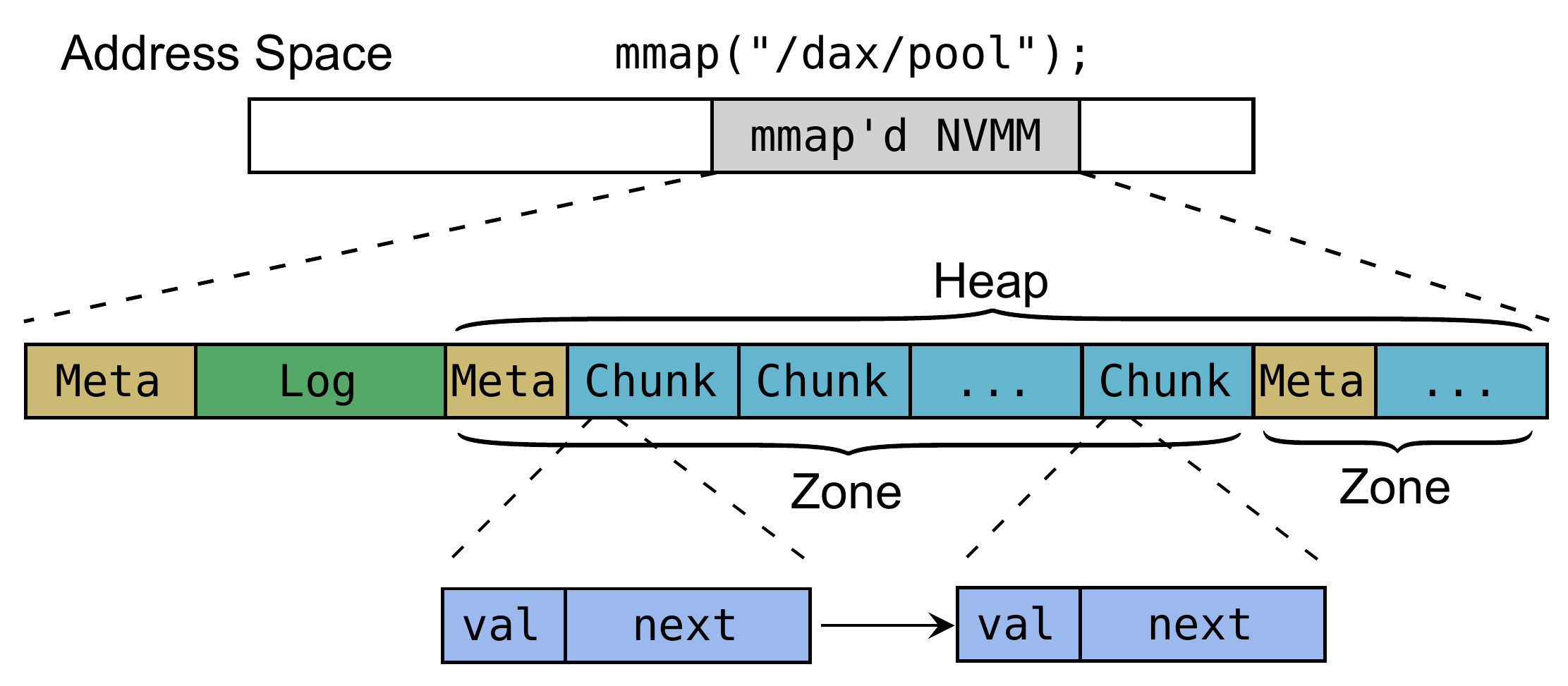,{
\figtitle{\DAXmmapd{} NVMM as an object store} \Lpo{} divides the mapped
         space into zones and chunks for memory management.
         %In this example, the root object stores a key-value pair.
         The \texttt{val} field is a 64-bit integer and the \texttt{next} field is a
         persistent pointer (\pmemoid{}) pointing to the next node object in the pool.
},fig:pmemobj]

In this section, we describe \lpo{}'s programming model. \Lpo{} is a
well-supported, open-source C library for programming with \DAXmmapd{} NVMM. It
provides facilities for memory management and software transactions that let
applications build a persistent object store.  \FTLib{}'s interface and implementation are based on 
\lpo{} from PMDK v1.5.

Linux exposes NVMM to the user-space as memory-mapped files
(\reffig{fig:pmemobj}). \Lpo{} (and \FTLib{}) refer to the mapped file as a
\emph{pool} of persistent objects. Each pool spans a continuous range of virtual
addresses.

\begin{lstlisting}[
  style=cstyle, numbers=left,
  keywordstyle={[2]{\color{blue}}}, label={lst:pmemobj},
  morekeywords={[2]{pmemobj_open, pmemobj_root, pmemobj_direct, pmemobj_persist,
                    pmemobj_alloc, pmemobj_tx_add_range, pmemobj_tx_alloc,
                    pmemobj_close, TX_BEGIN, TX_END, TX_ONABORT}},
  caption={A \lpo{} program - First modify a node value in a linked
          list, and later allocate and link a new node from the pool.
        }
]
PMEMobjpool *pool = pmemobj_open("/dax/pool");
...
struct node *n = pmemobj_direct(node_oid);
n->val = value;
pmemobj_persist(pool, &n->val, 8);
...
TX_BEGIN(pool) {
  n = pmemobj_direct(node_oid);
  pmemobj_tx_add_range(node_oid, 0, sizeof(*n));
  n->next = pmemobj_tx_alloc(...);
} TX_ONABORT {
  /* handling transaction aborts */
} TX_END
...
pmemobj_close(pool);
\end{lstlisting}

Within a pool, \lpo{} reserves a metadata region that contains information such
as the pool's identification (64-bit UUID) and the offset to a ``root object''
from which all other live objects are reachable.  Next, is an area reserved
for transaction logs.  \Lpo{} uses redo logging for its metadata updates and undo
logging for application object updates.  Transaction logs reside in one of two
locations depending on their sizes.  Small log entries live in the provisioned ``Log''
region, as shown in \reffig{fig:pmemobj}.
Large ones overflow into the ``Heap'' storage area.

The rest of the pool is the persistent heap. \Lpo{}'s NVMM allocator (a
persistent variant of \texttt{malloc}/\texttt{free}) manages it.  The
allocator divides the heap's space into several ``zones'' as shown in
\reffig{fig:pmemobj}.  A zone contains metadata and a sequence of ``chunks.''
The allocator divides up a chunk for small objects and coalesces
adjacent chunks for large objects. By default, a zone is 16~GB, and a chunk is
256~KB.

\reflst{lst:pmemobj} presents an example to highlight the key concepts of NVMM
programming. The code performs two independent operations on a persistent linked
list: one is to modify a node's value, and another is to allocate and link a new
node.

This example demonstrates two styles of crash-consistent NVMM programming:
\emph{atomic}-style (lines 3-5) for a simple modification that is 8 bytes or
smaller, and \emph{transactional}-style (lines 7-13) for arbitrary-sized NVMM
updates.

Building data structures in NVMM using \lpo{} (or any other persistent object
library) differs from conventional DRAM programming in several ways:

\boldparagraph{Memory Allocation} \Lpo{} provides crash-consistent NVMM
allocation and deallocation functions: \texttt{pmemobj\_tx\_alloc/pmemobj\_tx\_free}.
They let the programmer specify object type and size to allocate and prevent
orphaned regions in the case of poorly-time crashes.

\boldparagraph{Addressing Scheme} Persistent pointers within a pool must remain
valid regardless of at what virtual address the pool resides. \Lpo{} uses a
\pmemoid{} data structure to address an object within a pool. It consists of a
64-bit file ID and a 64-bit byte offset relative to the start of the file. The
\texttt{pmemobj\_direct()} function translates a \pmemoid{} into a native
pointer for use in load or store instructions.

\boldparagraph{Failure-atomic Updates} Modern x86 CPUs only guarantee that
8-byte, aligned stores atomically update NVMM~\cite{intel2017faq}. If
applications need larger atomic updates, they must manually construct software
transactions. \Lpo{} provides undo log-based transactions. The application
executes stores to NVMM between the \texttt{TX\_BEGIN} and \texttt{TX\_END}
macros, and snapshots (\texttt{pmemobj\_tx\_add\_range}) a range of object data
before modifying it in-place.

\boldparagraph{Persistence Ordering} Intel CPUs provide cache flush/write-back
(e.g., \texttt{CLFLUSH(OPT)} and \texttt{CLWB}) and memory ordering (e.g.,
\texttt{SFENCE}) instructions to make guarantees about when stores become
persistent. In \reflst{lst:pmemobj}, the \texttt{pmemobj\_persist} function and
\texttt{TX} macros integrate these
instructions to flush modified object ranges.

\Lpo{} supports a replicated mode that requires a replica pool, doubling the
storage the object store requires. \Lpo{} applies updates to both pools to
keep them synchronized.

Replicated \lpo{} can detect and recover from media errors only when the object
store is offline, and it cannot detect or recover from data corruption caused
by errant stores to NVMM -- so-called ``scribbles,'' that might result from a
buffer overrun or dereferencing a wild pointer.

\wfigure[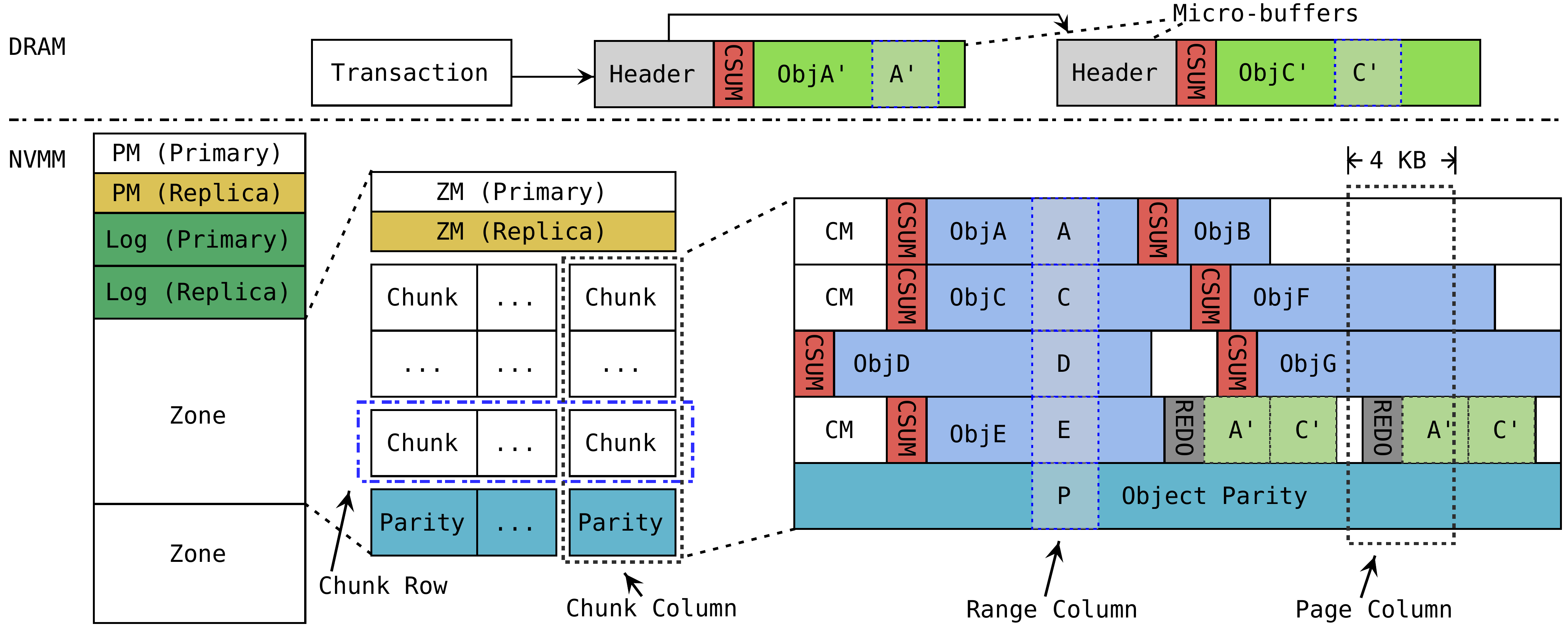,{
       \figtitle{Data protection scheme in \FTLib{}} \FTLib{} protects pool
         metadata (PM), zone metadata (ZM), and chunk metadata (CM).
         In the highlighted range column, $P = A \oplus C \oplus D \oplus E$.
         One thread's transaction is modifying ranges $A$ and $C$ of two
         objects. \FTLib{} keeps modified data in redo log entries (checksummed
         and replicated) when the transaction commits. The DRAM part shows
         \mbuf{}s for the two objects.
},fig:layout]

\section{\FTLib{} Design} \label{sec:design}

\FTLib{} allows programmers to build complex, crash-consistent persistent data
structures that are also robust in the face of media errors and software
``scribbles'' that corrupt data. \FTLib{} satisfies all of the criteria listed
in \refsec{sec:intro}. This section describes its architecture and highlights
the key challenges that \FTLib{} addresses to meet those requirements. In
particular, \FTLib{} provides the following features unseen in prior works.

\begin{itemize}

\item It provides fast, space-efficient recovery from media errors and scribbles.

\item It uses checksums to protect object integrity and supports incremental checksum updates.

\item It integrates parity and checksum updates into an NVMM transaction system.

\item It periodically scrubs data to identify corruption.

\item It detects and recovers from media errors and scribbles online.

\end{itemize}

\FTLib{} guarantees that it can recover from the loss of any single 4~KB page of
data in a pool.  In many cases, it can recover from the concurrent loss of multiple pages.

We begin by describing how \FTLib{} organizes data to protect user objects,
library metadata, and transaction logs using a combination of parity,
replication, and checksums. Next, we describe \mbufs{} and explain how they
allow \FTLib{} to preserve a simple programming interface and protect against software
scribbles.
Then, we explain how \FTLib{} detects and prevents NVMM corruption and
elaborate on \FTLib{}'s transaction implementation with support for efficient,
concurrent updates of object parity. Finally, we discuss
how \FTLib{} restores data integrity after corruption and crashes.

\subsection{\FTLib{}'s Data Organization} \label{sec:data}

\FTLib{} uses replication for its internal metadata and RAID-style parity for
user objects to provide redundancy for corruption recovery. The MCE mechanism
described in \refsec{sec:errmodel} and object checksums in \FTLib{} detect
corruption.

% illustrates a zone's chunks that span a continuous address range of NVMM.
\FTLib{} views a zone's chunks as a two-dimensional array
as shown in the middle of \reffig{fig:layout}. Each \emph{chunk row}
contains multiple, contiguous chunks and the chunks ``wrap around'' so that the
last chunk of a row and the first chunk of the next are adjacent. \FTLib{}
reserves the last chunk row for parity.

In our description of \FTLib{}, we define a \emph{page column} as a one
page-wide, aligned column that cuts across the rows of a zone. A \emph{range
column} is similar, but can be arbitrarily wide (no more than a chunk row's size).

Initializing a parity-coded NVMM pool requires zeroing out all the bytes in the
file. This is a one-time overhead when creating a pool file and does not affect
run-time performance. We report this latency in \refsec{sec:eval}.

To detect
corruption in user objects, \FTLib{} adds a 32-bit checksum to the object's
header.  The header also contains the object's size (64-bit)
and type (32-bit). The compiler determines type values according to user-defined
object types. \FTLib{} inherits this design from \lpo{} and changes the type
identifier from 64-bit to 32-bit for the checksum.

\FTLib{}'s object placement is independent of chunk and row boundaries. Objects
can be anywhere within a zone, and they can be of any size (up to the zone
size).

In addition to user objects, the library maintains metadata for the pool, zones,
and chunks, including allocation bitmaps.  \FTLib{}
checksums these data structures to detect corruption and replicates the pool's
and zones' metadata for fault tolerance. These structures are small (less than
0.1\% for pools larger than 1~GB), so replicating them is not expensive.
\FTLib{} uses zone parity to support recovery of chunk metadata.

\FTLib{} checksums transaction logs and replicates them for redundancy. It
treats log entries in zone storage as zeros during parity calculations. This
prevents parity update contention between log entries and user objects (see
\refsec{sec:update}).

\boldparagraph{Fault Tolerance Guarantees}
\FTLib{} can tolerate a single 4~KB media error anywhere in the pool, regardless
of whether it is a data page or a parity page. Based on the bad page's address
\FTLib{} can locate its page column and restore its data using other healthy
pages.

Faults affecting two pages of the same page column may cause data loss if the
corrupted ranges overlap. If an application demands more robust fault tolerance,
it can increase the chunk row size, reducing the number of rows and,
consequently, the likelihood that two corrupt pages overlap.

\FTLib{} can recover from scribbles (contiguous overwrites caused by software
errors) on NVMM data up to a chunk-row size. By default, \FTLib{} uses 100 chunk
rows, and parity consumes $\sim$1\% of a pool's size (e.g., 1~GB for a 100~GB pool).

\subsection{\Mbufing{} for NVMM Objects} \label{sec:mbuf}

\begin{table*} % put here for showing on the next page
\centering
\resizebox{\textwidth}{!}{
    \begin{tabular}{ll}
      \toprule
      Function & Semantics\\
      \midrule
      \texttt{pgl\_tx\_begin()/commit()/end(), etc.}
            & Control the lifetime of a \FTLib{} transaction.\\
      \texttt{pgl\_tx\_alloc()/free()}
            & Allocate or deallocate an NVMM object.\\
      \texttt{pgl\_tx\_open(PMEMoid oid, ...)}
            & Create a thread-local \mbuf{} for an NVMM object. Verify (and restore)\\
            & the object integrity, and return a pointer to the \mbufed{} user object.\\
      \texttt{pgl\_tx\_add\_range(PMEMoid oid, ...)}
            & Invoke \texttt{pgl\_tx\_open} and then mark a range of the object that will be modified.\\
      \texttt{pgl\_get(PMEMoid oid)}
            & Get access to an object, either directly in NVMM or in its \mbuf{},\\
            & depending on the transaction context. By default, it does not verify the checksum.\\
      \midrule
      \texttt{pgl\_open(PMEMoid oid, ...)}
            & Create a \mbuf{} for an NVMM object without a transaction. Check the\\
            & object integrity, and return a pointer to the \mbufed{} user object.\\
      \texttt{pgl\_commit(void *uobj)}
            & Automatically start a transaction and commit the modified user object in a\\
            & \mbuf{} to NVMM.\\
      \bottomrule
    \end{tabular}
}
\caption{
  The \FTLib{} API - \FTLib{}'s interface mirrors \lpo{}'s except that \FTLib{}
  does not allow direct writing to NVMM. \FTLib{} provides single-object
  transactions using \texttt{pgl\_open} and \texttt{pgl\_commit} to convert
  application code using \lpo{}'s atomic updates.
}
\label{tab:api}
\end{table*}

\FTLib{} introduces \mbufing{} to hide the complexity of updating checksums and
parity when modifying NVMM objects. Adding checksums to objects and protecting
them with parity makes updates more complex, since all three -- object data,
checksum, and parity -- must change at once to preserve consistency.  This
challenge is especially acute for the atomic programming model as shown in
\reflst{lst:pmemobj} (line 3-5) because a single 8-byte NVMM write cannot host
all these updates.

\Mbuf{}ing creates a shadow copy of an NVMM object in DRAM, which separates an
object's transient and persistent versions (\reffig{fig:layout}).
In \reflst{lst:ftlib}, \texttt{pgl\_open} creates a \mbuf{} for the node
object by allocating a DRAM buffer and copying the node's data from NVMM.
It also verifies the object's checksum and performs corruption recovery if
necessary.

The application can modify the \mbufed{} object without concern for its checksum,
parity, and crash-consistency because changes exist only in the \mbuf{}.
When the updates finish, \texttt{pgl\_commit} starts a
transaction that atomically updates the NVMM object, its checksum, and parity
(described below). Compared to line 3-5 of
\reflst{lst:pmemobj}, \FTLib{} retains the simple, atomic-style programming
model for modifying a single NVMM object, and it supports updates within an
object beyond 8 bytes.

Each \mbuf{}'s header contains information such as its
NVMM address, modified ranges, and status flags (e.g., allocated or modified).
We elaborate on \FTLib{}'s programming interface and how to construct complex
transactions with \mbufing{} in \refsec{sec:tx}.

\begin{lstlisting}[
  style=cstyle, numbers=left,
  label={lst:ftlib},
  keywordstyle={[2]{\bfseries\color{teal}}},
  morekeywords={[2]{pgl_open, pgl_commit, pgl_tx_begin, pgl_tx_alloc,
                    pgl_tx_open, pgl_tx_add_range, pgl_tx_commit, pgl_tx_end,
                    PGL_TX_BEGIN, PGL_TX_ONABORT, PGL_TX_END}},
  caption={A \FTLib{} transaction for a single-object - This snippet corresponds to line
           3-5 of \reflst{lst:pmemobj}.
           }
]
struct node *n = pgl_open(node_oid);
n->val = value;
pgl_commit(pool, n);
\end{lstlisting}

Another important consideration for \mbufing{} is to prevent misbehaving
software from corrupting NVMM. If an application's code can directly write to
NVMM, as \lpo{} allows to, software bugs such as buffer overflows and using
dangling pointers can easily cause NVMM corruption. Conventional debugging tools
for memory safety, such as Valgrind \cite{valgrind} and AddressSanitizer
\cite{asan}, insert inaccessible bytes between objects as ``redzones'' to trap
illegal accesses.
This approach fails to work for directly accessed NVMM objects because once they
are allocated, there is no guarantee for spacing between them, and thus,
redzones may land on a nearby, accessible object. One viable approach to using
these tools is to let the NVMM allocator insert redzones. However, the presence
of redzone bytes will pollute the pool and may exacerbate fragmentation.

Using \mbufs{} isolates transient writes from persistent data, and since
\mbuf{}s are dynamically allocated using \texttt{malloc()}, they are compatible
with existing memory debugging tools. Without using debugging tools, \FTLib{}
also protects \mbuf{}s by inserting a 64-bit ``canary'' word in each \mbuf{}'s
header and checks its integrity before writing back to NVMM. On transaction
commit, if \FTLib{} detects a canary mismatch, it aborts the transaction to
avoid propagating the corruption to NVMM. Pangolin uses checksums to detect
corruptions that may bypass the canary protection.

\subsection{Detecting NVMM Corruption} \label{sec:detecting}

\FTLib{} uses three mechanisms to detect NVMM corruption. First, it installs a
handler for \sigbus{} (see \refsec{sec:errmodel}) that fires when the Linux
kernel receives an MCE.  A signal handler has access to the address the offending
load accessed, and \FTLib{} can determine what kind of data (i.e., metadata or a
user object) lives there and recover appropriately.  This mechanism detects
media failures, but it cannot discover corrupted data caused by software
``scribbles.''

To detect scribbles, \FTLib{} verifies the integrity of user objects using their checksums.
Verifying checksums on every access can be expensive. To limit this cost, by
default \FTLib{} only verifies checksums during \mbuf{} creation before any
object is modified in a transaction. This keeps \FTLib{} from recalculating a
new checksum based on corrupt data. For read-only objects that are accessed by
\texttt{pgl\_get} without \mbufing{}, by default \FTLib{} does not verify
checksums. To protect them, \FTLib{} provides two alternative operation modes:
``Scrub'' mode runs a scrubbing thread that verifies and restores the
whole pool's data integrity when a preset number of transactions have
completed, and ``Conservative'' mode verifies the checksum for every object
access (including \texttt{pgl\_get}). We evaluate the impact of different
checksum verification policies in \refsec{sec:eval}.

Finally, Linux keeps track of known bad pages of NVMM across reboots. When
opening a pool or during its scrubbing, \FTLib{} can extract this information
and recover the data in the reported pages (not currently implemented).

\ignore{
  Both methods only perform checksum verification when there are no active
  transactions. In the future work, we also plan to provide a scrubbing function
  that executes periodically to verify object checksums, according to practical
  NVMM error statistics.

  Another way is that the error arises when an application is reading its
  affected memory range. In this case, Linux sends a \sigbus{} to the reading
  thread, and the thread's normal execution flow is interrupted. \FTLib{}
  implements a specialized signal handler to intercept this event, avoiding
  crashing the application.

  In the signal handler \FTLib{} can learn about the faulty page address and
  repair the corrupted data. After returning from the signal handler, the
  program can resume execution from where it was interrupted. The process is
  conceptually similar to handle a page fault in the user-space.
}

\subsection{Fault-Tolerant Transactions} \label{sec:tx}

Failure-atomic transactions are central to \FTLib{}'s interface, and they must
include verification of data integrity and updates to the checksums and parity
data that protect objects. \reftab{tab:api} summarizes \FTLib{}'s core
functions.

\FTLib{} supports arbitrary-sized transactions and we have
made similar APIs and macros as \lpo{}'s. The program in \reflst{lst:pmemobj}
can be easily transformed to \FTLib{} using
equivalent functions. One subtle difference is in the handling of atomic-style
updates, as shown in \reflst{lst:ftlib}.

In \FTLib{}, each thread can execute one transaction or nested transactions
(same as \lpo{}). Concurrent transactions can execute if each one is associated
with a different thread.
Currently, \FTLib{} does not allow concurrent transactions to modify the same
NVMM object. Concurrently modifying a shared object may cause data inconsistency
if one transaction has to abort. \Lpo{} has the same limitation~\cite{pmemcheck}.

Each transaction manages its own \mbuf{}s using a thread-local hashmap~\cite{cuckoo},
indexed by an NVMM object's \pmemoid{}. Therefore, in
a transaction, calling \texttt{pgl\_tx\_open} for the same object either creates
or retrieves its \mbuf{}. Multiple \mbuf{}s opened in one transaction form a linked
list as shown in \reffig{fig:layout}.
\Mbuf{}s for one transaction are not visible in other transactions, providing isolation.

If a transaction modifies an object, \FTLib{} copies it to a \mbuf{},
performs the changes there, and then propagates the changes to NVMM during
commit.  Since changes occur in DRAM (which does not require undo information),
\FTLib{} implements redo logging.

At transaction commit, \FTLib{} recomputes the checksums for modified \mbufs{},
creates and replicates redo log entries for the modified parts of the \mbufs{}
and writes these ranges back to NVMM objects.
Then, it updates the
affected parity bits (see \refsec{sec:update}) and marks the transaction
committed. Finally, \FTLib{} garbage-collects its logs and closes thread-local
\mbuf{}s.

If a transaction aborts, either due to unrecoverable data corruption or other
run-time errors, \FTLib{} discards the transaction's \mbuf{}s without touching
NVMM.

A transaction can also allocate and deallocate objects. \FTLib{} uses redo
logging to record NVMM allocation and free operations, just as \lpo{} does.

For read-only workloads, repeatedly creating \mbuf{}s and verifying object
checksums can be very expensive.  Therefore, \FTLib{} provides
\texttt{pgl\_get} to gain direct access to an NVMM object without verifying the
object's checksum.  The application can verify an object's integrity manually
as needed or rely on \FTLib{}'s periodic scrubbing mechanism.  Inside a
transaction context, \texttt{pgl\_get} returns a pointer to the object's
\mbuf{} to preserve isolation.

\subsection{Parity and Checksum Updates} \label{sec:update}

Objects in different rows can share the same range of parity, and we say these objects
\emph{overlap}.  Object overlap leads to a challenge for updating the shared
parity because updates from different transactions must serialize but naively
locking the whole parity region sacrifices scalability.

For instance, using $ObjA$ and $ObjC$ in \reffig{fig:layout}, suppose two
different transactions modify them, replacing $A$ with $A'$ and
$C$ with $C'$, respectively. After both transactions update $P$, the parity
should have the value $P' = A' \oplus C' \oplus D \oplus E$ regardless of how
the two transaction commits interleave.

\FTLib{} uses a combination of two techniques that exploit the commutativity of
XOR and fine-grained locking to preserve correctness and scalability.

\boldparagraph{Atomic parity updates} \label{sec:atomic} The first approach uses
the atomic XOR instruction (analogous to an atomic increment) that modern
CPUs provide to perform incremental parity updates for changes to each
overlapping object.

In our example, we can compute two parity patches:
$\Delta{}A = A \oplus A', ~ \Delta{}C = C \oplus C'$ and then rewrite
$P'$ as $P \oplus \Delta{}A \oplus \Delta{}C$. Since XOR commutes and is a
bit-wise operation, the two threads can perform their updates without
synchronization.

\ignore{
  Because parity words are independent of each other, \FTLib{} can perform an
  efficient delta-based parity update only using modified ranges of objects. A
  \emph{range delta} is defined as the XOR result between the old data and new
  data of a modified object range, i.e., , and \FTLib{} can derive $P'_1$ using
  range deltas as $P' = P \oplus \Delta{}A \oplus \Delta{}C$, eliminating $D$
  and $E$ from the computation.

  Since XOR computations are commutative, $\Delta{}A$ and $\Delta{}C$ can apply
  to $P$ in any order. We can leverage the byte-addressability of NVMM and CPU's
  atomic XOR instructions to perform a lock-free parity update. This is also
  correct even if the size of $\Delta{}A$ and $\Delta{}C$ are not equal or not
  aligned. Once the parity updates complete, \FTLib{} flushes the modified
  parity range to ensure the changes are persistent.
}

\boldparagraph{Hybrid parity updates} \label{sec:hybrid} Atomic XOR is slower
than normal or vectorized XOR. For small updates, the latency difference
between them is not significant, and \FTLib{} prefers atomic XOR instructions
to update parity without the need for locks.
But for large parity updates, atomic XOR can be inefficient.
% \reffig{fig:xor-latency} illustrates the trade-off.
Therefore, \FTLib{}'s hybrid parity scheme switches to vectorized XOR for large
transfers.

To coordinate large and small parity updates, \FTLib{} uses
\emph{parity range-locks}, that work similarly as reader/writer locks (or shared mutex):
Small writes take shared ownership of a range lock and update parity with
atomic XOR instructions. Large updates using vectorized XORs take
exclusive ownership of a range-lock, and only one thread can modify parity in a
locked range. If one update involves multiple range-locks, serialization happens
on a per-range-lock basis.

The managed size of a parity range-lock depends on the performance trade-off
between \FTLib{}'s parity mode and \lpo{}'s replication mode. We discuss this
in \refsec{sec:eval}.

\FTLib{} refreshes an object's checksum in its \mbuf{} before updating parity,
and it considers the checksum field as one of the modified ranges of the object.
Checksums like CRC32 requires recomputing the checksum using the whole object.
This can become costly with large objects. Thus, \FTLib{} uses
Adler32~\cite{waflmetaprot}, a checksum that allows incremental updates, to
make the cost of updating an object's checksum proportional to the size of the
modified range rather than the object size.

We implement \FTLib{}'s parity and checksum updates using the Intelligent
Storage Acceleration Library (ISA-L) \cite{isa-l}, which leverages SIMD
instructions of modern CPUs for these data-intensive tasks.

\boldparagraph{Protections for other transaction systems} Other NVMM persistent
object systems could apply \FTLib{}'s techniques for parity and checksum
updates.  For example, consider an undo logging (as opposed to \FTLib{}'s redo logging) system that first
stores a ``snapshot'' copy of an object in the log before modifying the original in-place.
In this case, the system could compute a
parity patch using the XOR result between the logged data (old) and the
object's data (new). Then, it can apply the parity patch using the hybrid
method we described in this section.

\ignore{
  updates, \FTLib{} can compute a new checksum value for each modified range of
  an object. The incremental updating algorithm~\cite{waflmetaprot} we use is
  conceptually similar to the delta-based parity update, but the delta value is
  the 2s-compliment difference, i.e., $|A - A'|$ or $|C - C'|$.

  A checksum value is associated with each object, and it does not introduce any
  concurrency challenges to compute and store. \FTLib{} modifies its
  corresponding parity bits using hybrid parity updating scheme.
}

\subsection{Recovering from Faults} \label{sec:recoverfaults}

In this section, we discuss how \FTLib{} recovers data integrity from both
NVMM corruption and system crashes.

\boldparagraph{Corruption recovery}
\FTLib{} uses the same algorithm to recover from errors regardless of how it
detects them (i.e., via \sigbus{} or a checksum mismatch).

The first step is to pause the current thread's transaction, and to wait until
all other outstanding transactions have completed. Meanwhile, \FTLib{}
prevents the initialization of new transactions by setting the pool's ``freeze''
flag. This is necessary because, during transaction committing, parity data may be
inconsistent.

Once the pool is frozen, \FTLib{} uses the parity bits and the corresponding
parts of each row in the page column to recover the missing data.

\FTLib{} preserves crash-consistency during repair by making persistent records
of the bad pages under recovery. Recovery is idempotent, so it can simply
re-execute after a crash.

\FTLib{}'s current implementation only allows one thread to perform any online
corruption recovery, and if the thread is executing a transaction, online
recovery only works if the thread has not started committing. If
two threads encounter faults simultaneously, \FTLib{} kills
the application and performs post-crash recovery (see below) when it restarts.
Supporting multi-threaded online recovery, and allowing it to work when threads
have partially written NVMM is possible, but it requires complex
reasoning about how to restore the data and its parity to a
consistent state.

\boldparagraph{Crash recovery}
\FTLib{} handles recovery from a crash using its redo logs. It must also protect
against the possibility that the crash occurred during a parity update.

To commit a transaction, \FTLib{} first ensures its redo logs are persistent and
replicated, and then updates
the NVMM objects and their parity. If a crash happens before redo logs
are complete, \FTLib{} discards the redo logs on reboot without touching the objects
or parity. If redo logs exist, \FTLib{} replays them to update the objects and
then recomputes any affected parity ranges using
the data written during replay (which is now known to be correct) and the data
from the other rows.

\FTLib{} does not log parity updates because it would double the cost of
logging. This does raise the possibility of data loss if a crash occurs during
a parity update and a media error then corrupts data of the same page column
before recovery can complete. This scenario requires the simultaneous loss of
two pages in the same page column due to corruption and a crash, which we expect to be rare.

\section{Evaluation} \label{sec:eval}

In this section, we evaluate \FTLib{}'s performance and the overheads it incurs
by comparing it to normal \lpo{} and its replicated version. We start with our
experimental setup and then consider its storage requirements, latency impact,
scalability, application-level performance, and corruption recovery.

\wfigure[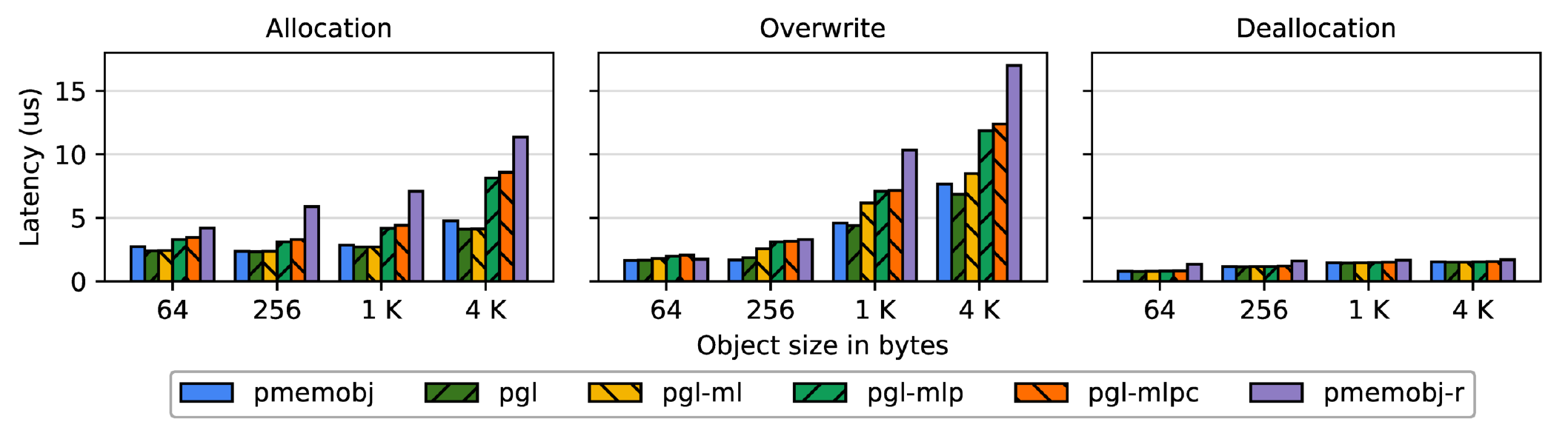,{
         \figtitle{Transaction performance}
         Each transaction allocates, overwrites, or frees one object of varying
         sizes. \FTLib{}'s latencies are similar to \Po{}'s. \FTLib{}-MLP is
         mostly better than \Por{} because updating parity using atomic XOR and \clwb{}
         instructions is faster than mirroring data in a separate file.
         },fig:latency]

\subsection{Evaluation Setup}

We perform our evaluation on a dual-socket platform with Intel's Optane DC
Persistent Memory~\cite{optane-pmm}. The CPUs are 24-core engineering samples of
the Cascade Lake generation. Each socket has 192~GB DDR4 DRAM and 1.5 TB NVMM.
We configure the persistent memory modules in \emph{AppDirect} mode and run
experiments on one socket using its local DRAM and NVMM. A recent
report~\cite{izraelevitz2019basic} studying this platform provides more
architectural details. % and basic performance measurements.

The CPU provides the \clwb{} instruction for writing-back cache lines to NVMM,
non-temporal store instructions to bypass caches, and the \sfence{} instruction
to ensure persistence and memory ordering. It also has atomic XOR and AVX
instructions that our parity and checksum computations use.

The evaluation machine runs Fedora 27 with a Linux kernel version 4.13 built
from source with the NOVA~\cite{nova} file system. We run experiments with both
Ext4-DAX~\cite{ext4dax} and NOVA, and applications use \texttt{mmap()} to access
NVMM-resident files. The performance is similar on the two file systems because
DAX-\texttt{mmap()} essentially bypasses them.

On our evaluation machine, we found that updating parity with atomic XORs
becomes worse than \lpo{}'s replication mode when the modified parity range is
greater than 8~KB, so we set 8~KB as the threshold to switch between those
parity calculation strategies (see \refsec{sec:hybrid}).

\begin{table}
\centering
  \resizebox{\columnwidth}{!} {
    \begin{tabular}{|l|l|}\hline
    % \toprule
       \Po{}         & \texttt{libpmemobj} baseline from PMDK v1.5 \\\hline
       \FTLib{}      & \FTLib{} baseline w/ \mbufing{} only \\\hline
    %  \FTLib{}-M    & \FTLib{} w/ metadata replication \\\hline
       \FTLib{}-ML   & \FTLib{} + metadata and redo log replication \\\hline
       \FTLib{}-MLP  & \FTLib{}-ML + object parity\\\hline
       \FTLib{}-MLPC & \FTLib{}-MLP + object checksums\\\hline
       \Por{}        & \texttt{libpmemobj} w/ one replication in another file \\\hline
    % \bottomrule
    \end{tabular}
  }
  \caption{Library operation modes for evaluation - In the figures, we
  abbreviate \FTLib{} as \texttt{pgl}.}
\label{tab:modes}
\end{table}

\reftab{tab:modes} describes the operation modes for our evaluations. The
\FTLib{} baseline implements transactions with \mbufing{}. It uses buffer
canaries to prevent corruption from affecting NVMM, but it does not have parity
or checksum for NVMM data.

We evaluate versions of \FTLib{} that incrementally add metadata and log
replication (``+ML''), object parity (``+MLP''), and checksums (``+MLPC'').
We combine the impact of metadata updates with log replication because
metadata updates are small and cheap in our evaluation.

\Por{} is the replication mode of \lpo{} that mirrors updates to a replica pool
during transaction commit.  Comparing \FTLib{}-MLP and \Por{} is especially
useful because the two configurations protect against the same types of data corruption:  media errors but not
scribbles.

\subsection{Memory Requirements}

We discuss and evaluate \FTLib{}'s memory requirements for both NVMM and DRAM.

\boldparagraph{NVMM}
All our \FTLib{} experiments use a single pool of 100~GB that contains 6$\times$
16~GB zones. \FTLib{} replicates all the pool's metadata in the same file, which
occupies a fixed $\sim$20~MB. The rest of the space is for user objects and
their protection data. By default, \FTLib{} uses 100 chunk rows, so each zone
has about 160~MB parity, and that totally occupies $\sim$1\% of the pool's
capacity. \Por{} uses a second 100~GB file as the replica, doubling the cost of
NVMM space requirement.

When using parity, \FTLib{} has to zero out the whole pool to ensure all
zones are initially parity-coded. This takes about 130 seconds. It is a
one-time overhead for creating the pool and excluded from the following
evaluations.

\boldparagraph{DRAM}
\FTLib{} uses \texttt{malloc()}'d DRAM to construct \mbufs{}. The required
DRAM space is proportional to ongoing transaction sizes. \reftab{tab:tx-sizes}
summarized the transaction sizes for the evaluated key-value store data
structures. \FTLib{} automatically recycles them on transaction commits.
In our evaluation experiments, \mbufing{} never exceeds using 50~MB of DRAM.

\subsection{Transaction Performance}

\wfigure[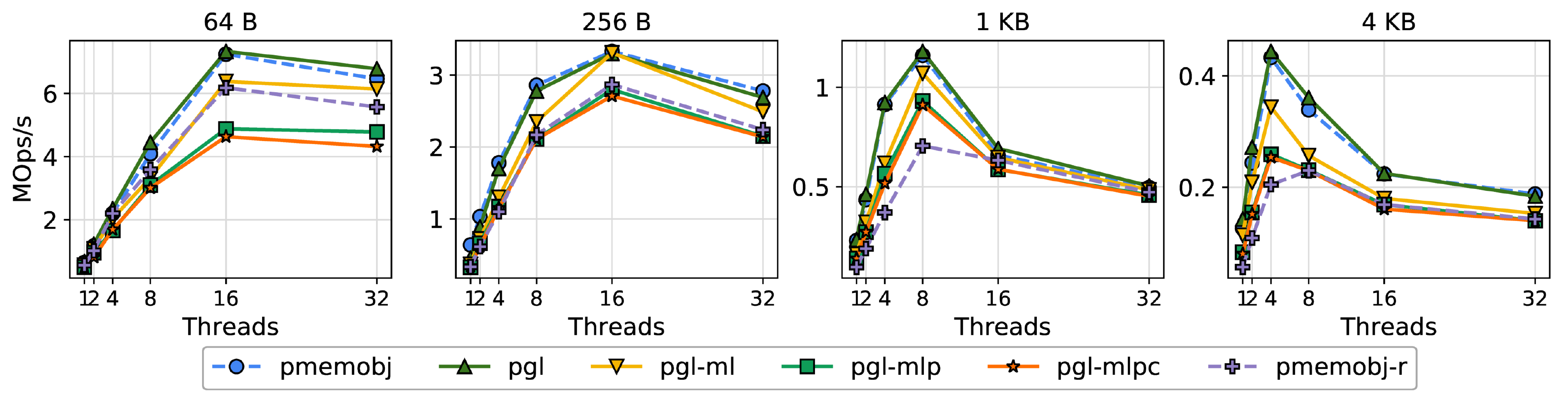,{
         \figtitle{Scalability}
         Concurrent workloads randomly overwrite objects of varying sizes.
         \FTLib{}-MLP scales as well as \Por{} or better for objects larger
         than 64~B. For 64~B objects, \FTLib{}-MLP is worse than \Por{} by between
         6\% and 25\% due to synchronization overhead for online recovery.
         },fig:scaling]

\reffig{fig:latency} illustrates the transaction latencies for three basic
operations on an NVMM object store: object allocation, overwrite, and
deallocation. Each transaction operates on one object, and we vary the size of
the object.

For allocation, latency grows with object size for all five configurations, due
to constructing the object and cache line write-back latency. \FTLib{} incurs
2\% - 13\% lower latencies than \Po{} due to its use of non-temporal stores for
write backs. An allocation operation does not involve object logging, so
\FTLib{}-ML shows performance similar to \FTLib{}. \FTLib{}-MLP adds overhead to update the
parity data. It outperforms \Por{} by between 1.2$\times$ and 1.9$\times$.
We found this is because updating parity using atomic XORs and \clwb{}s
incurs less latency than mirroring data in a separate file, as \Por{} does.

Adding checksum (\FTLib{}-MLPC) incurs less than 7\% overhead compared to
\FTLib{}-MLP.
Parity's impact is larger than checksum's
because updating a parity range demands values from three parts: the \mbuf{},
the NVMM object, and the old parity data, while computing a checksum only needs
data in a DRAM-based \mbuf{}. Moreover, \FTLib{} needs to flush the modified parity range
to persistence, which is the same size as the object. In contrary, updating
a checksum only writes back a single cache line that contains the checksum
value.

Overwriting an NVMM object involves transaction logging for crash consistency.
\FTLib{} and \Po{} store the same amount of logging data in NVMM, although they
use redo logging and undo logging for this purpose, respectively. Since log
entry size is proportional to an object's modified size, which is the whole
object in this evaluation, this cost grows with the object. With \FTLib{}, log
replication accounts for between 7\% to 25\% of the latency. Parity updates consume
between 8\% to 27\% of the extra latency, depending on object size, and
checksum updates account for less than 5\%. \FTLib{}-MLP's performance
for overwrites is 12\% worse than \Por{} for 64~B object updates and is between
1.1$\times$ and 1.5$\times$ better than \Por{} for objects larger than 64~B.

Deallocation transactions only modify metadata, so their latencies do not change
much.

\subsection{Scalability} \label{sec:scaling}

\begin{table*}
\centering
% \resizebox{\columnwidth}{!}{
  % \resizebox{.8\textwidth}{!}{
    \begin{tabular}{rlrrrrrr}\hline
    % \toprule
             &      & ctree          & rbtree        & btree        & skiplist     & rtree        &    hashmap  \\\hline
    % \midrule
    \multicolumn{2}{r}
      {Object size} &    56          &     80        &   304        &    408       &  4136        &  10~M (table), 40 (entry) \\\hline
    % \midrule
      \multirow{2}{*}{Insert}
                & New &    56 (1.00) &     80 (1.00) &  65.9 (0.22) &   408 (1.00) &  4502 (1.09) &  60.9 (1.00) \\
                & Mod & 127.6 (3.28) &  330.2 (5.13) & 381.2 (1.47) &  33.9 (2.50) & 200.0 (5.05) & 331.1 (4.21) \\\hline
      \multirow{2}{*}{Remove}
                & New &     0        &     0         &     0        &     0        & 184.1 (0.05) & 10.5 (\num{1e-5}) \\
                & Mod &  28.0 (0.50) &  202.8 (2.65) & 268.3 (0.90) &  16.9 (0.75) &  98.6 (2.52) &      254.3 (2.16) \\\hline
    % \bottomrule
    \end{tabular}
  % }
\caption{Data structure and transaction sizes - ``Insert'' and ``Remove'' show
         average transaction sizes for insertions and removals, respectively.
         ``New'' and ``Mod'' indicate average allocated
         and modified sizes. Value in parentheses is the average number of
         objects involved in the transaction.
        }
\label{tab:tx-sizes}
\end{table*}

\reffig{fig:scaling} measures \FTLib{}'s scalability by randomly
overwriting existing NVMM objects and varying the object sizes and the number of
concurrent threads.

\FTLib{} uses reader/writer locks to implement the hybrid parity update scheme
described in \refsec{sec:update}. The number of rows in a zone and the zone
size determine the granularity of these locks: For a fixed zone size, more rows
means fewer columns and fewer parity range-locks.

There is no lock contention in the results because the transactions use atomic XOR
instructions and can execute concurrently (only taking the reader locks).
Our configuration with 1\% parity (160~MB parity per 16~GB
zone) has 20~K range-locks per zone, so the chance of lock contention is slim
even with large updates (more than 8~KB) and many cores.

The graphs also show how each \FTLib{}'s fault-tolerance mechanisms affect
performance. \FTLib{}'s throughput is very close to \Po{}. \FTLib{}-MLP
mostly outperforms \Por{} for object updates that are 256~B or larger, up
to 1.5$\times$. But for 64~B object updates, it
performs worse than \Por{} by between 6\% and 25\%.
This is because when enabling parity, every \FTLib{} transaction checks the
pool freeze flag (an atomic variable), incurring
synchronization overhead. This overhead is noticeable for short transactions
with 64~B objects but becomes negligible for larger updates.
\FTLib{}-MLPC only performs marginally worse than \FTLib{}-MLP.

Scaling degrades for all configuration as update size and thread count grow
because the sustainable bandwidth of the persistent memory modules becomes saturated.

\subsection{Impacts on NVMM Applications}

\wfigure[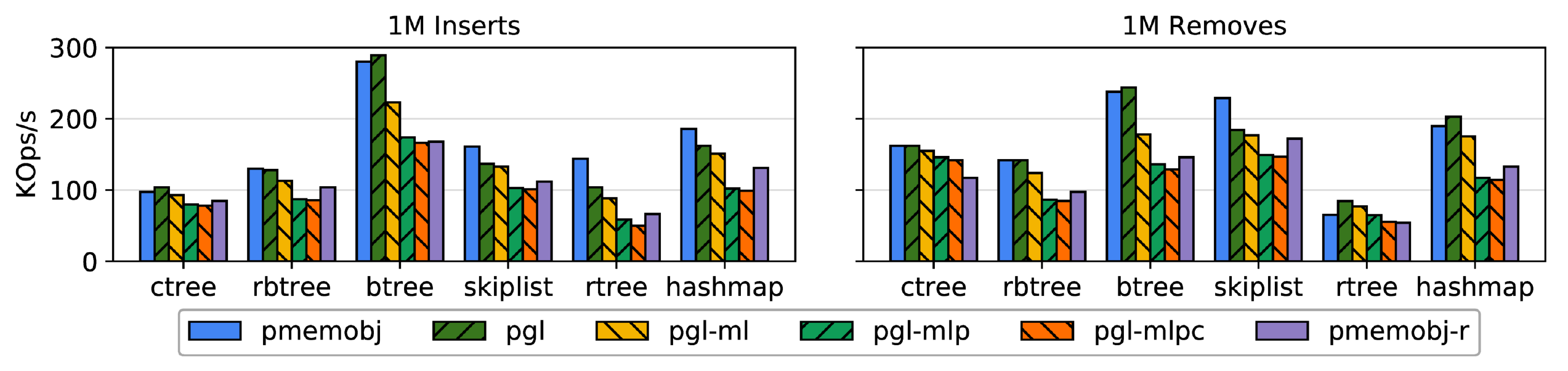,{
         \figtitle{Key-value store performance}
         Each transaction either inserts or removes one key-value pair from the
         data store. \FTLib{} performs similarly to \Po{} except for cases when a
         transaction's modified size is much less than an object's size
         (e.g., skiplist and rtree) due to \mbuf{}ing overhead.
         \FTLib{}-MLP's performance is 95\% of \Por{} on average.
         },fig:mapbench]

\cfigure[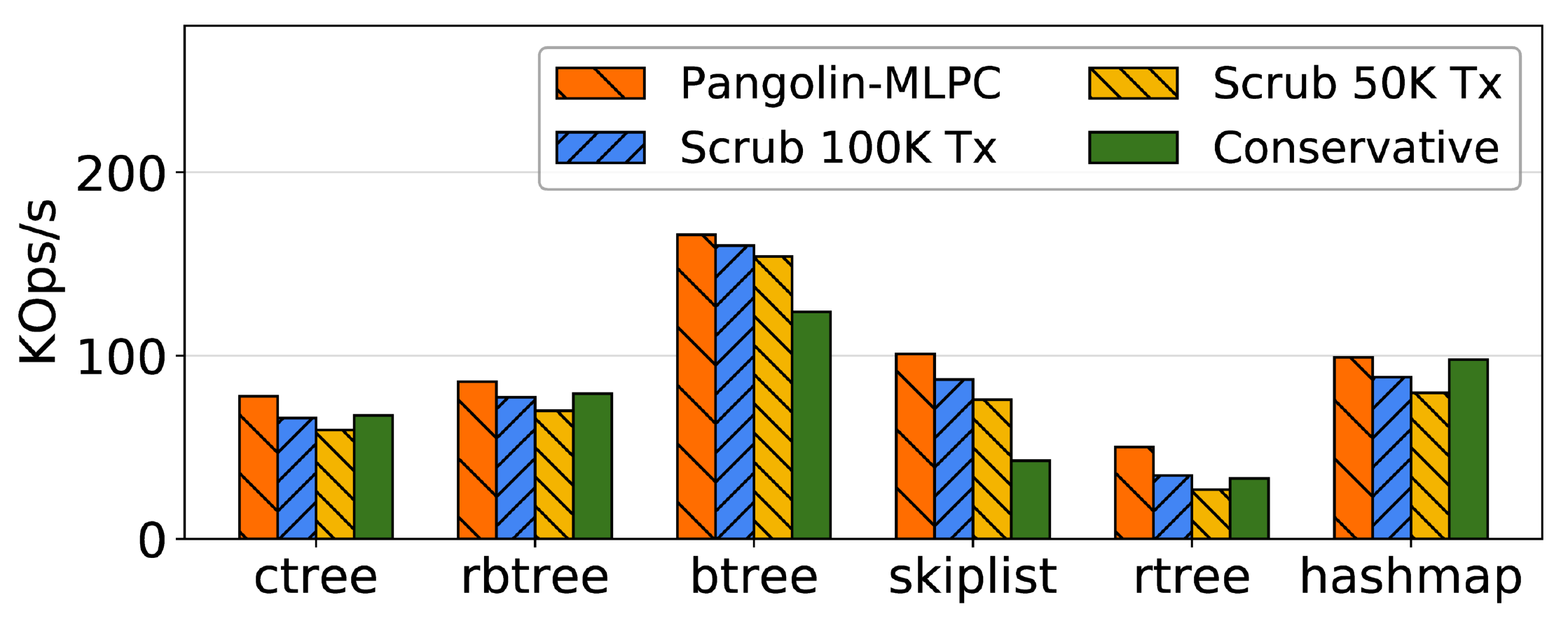,{
         \figtitle{Checksum verification impact} % For each data structure
           Pangolin-MLPC bars are the same as those in \reffig{fig:mapbench}
           for 1M Inserts. The cost of different policies depends strongly on
           data structures.
         },fig:checksum]

To evaluate \FTLib{} in more complex applications, we use six data structures
included in the PMDK toolkit: crit-bit tree (ctree), red-black tree (rbtree),
btree, skiplist, radix tree (rtree), and hashmap. They have a wide range of
object sizes and use a diverse set of algorithms to insert, remove, and lookup
values. We rewrite these benchmarks with \FTLib{}'s programming interface as
described in \refsec{sec:tx}.

\reftab{tab:tx-sizes} summarizes the object and transaction sizes for each workload.
The tree structures and the skiplist have a single type of object, which is the
tree or list node. Hashmap has two kinds of objects. One is the hash table
that contains pointers to buckets. The hash table grows as the application
inserts more key-value pairs. Each bucket is a linked list of fixed-sized entry
objects.

Each insertion or removal is a transaction processing a key-value pair.
The workloads involve a mix of object allocations, overwrites, and deallocations.
\reftab{tab:tx-sizes} shows, on average,
the number of bytes and objects (in parentheses) involved in each data
structure's transaction.
Deallocated sizes are not shown because they marginally affect the performance
differences (see \reffig{fig:latency}).

An average allocation size (``New'' rows in the table) smaller than the object
size means the data structure does not allocate a new object for every insert
operation (e.g., btree). The average modified sizes (``Mod'' rows) determine
the logging size and affect the performance drop between \FTLib{} and
\FTLib{}-ML. Note that a transaction does not necessarily modify (and log)
the whole range of an involved object.
The performance difference between \FTLib{}-ML and \FTLib{}-MLP
is a consequence of both allocated and modified sizes.

For insert transactions, \FTLib{} is faster than \Po{} for ctree and btree, but
slower than \Po{} for other data structures. This is because the slower
applications have relatively small modified sizes compared to the object sizes, and
\FTLib{}'s data movement from NVMM to \mbuf{}s overshadows its advantage for
whole-object updates, as shown in \reffig{fig:latency}.
For remove transactions, \FTLib{} is marginally faster than \Po{} except for the
case of skiplist, which is also because of the data movement caused by
\mbufing{}.

\FTLib{}-MLP's performance is 95\% of \Por{} on average, and it saves orders of
magnitude NVMM space by using parity data as redundancy. \FTLib{}-MLPC adds
scribble detection and performance drops by between 1.5\% to 15\% relative to
\FTLib{}-MLP. Adding object checksums impacts rtree's transactions the most because
the allocated object size is large, which requires more checksum computing time.

\FTLib{} does not impact the lookup performance because it performs direct NVMM
reads without constantly verifying object checksums. \FTLib{} ensures data
integrity with its checksum verification policy, as discussed in
\refsec{sec:detecting}.

\begin{table}
\centering
  \resizebox{\columnwidth}{!}{
    \begin{tabular}{lrrrrrr}\hline
    % \toprule
                     & ctree & rbtree & btree & sklist & rtree  & hmap \\\hline
    % \midrule
             Pmemobj &   1.0 &    1.0 &   1.0 &    1.0 &    1.0 &   1.0 \\
            Pgl-MLPC &  0.92 &   0.84 &  0.87 &   0.96 &   0.42 &  0.42 \\
          Scrub 100K &  0.10 &   0.09 &  0.09 &   0.10 &   0.04 &  0.05 \\
           Scrub 50K &  0.05 &   0.04 &  0.05 &   0.05 &   0.02 &  0.02 \\
        Conservative &     0 &      0 &     0 &     0  &     0  &     0 \\\hline
    % \bottomrule
    \end{tabular}
  }
  \caption{Vulnerability evaluation -  Each row shows object bytes (normalized
           to \Po{}) accessed without checksum verification.}
\label{tab:ftmodel}
\end{table}

\reffig{fig:checksum} illustrates the impact of different strategies for checksum
verification. We compare \FTLib{}'s default mode (\FTLib{}-MLPC) with two
``Scrub'' modes and a ``Conservative'' mode. The default mode only verifies
checksums for micro-buffered objects. In ``Scrub'' mode, a scrubbing thread
verifies data integrity of the whole object pool when a preset number (indicated
by legends in \reffig{fig:checksum}) of
transactions have completed. The ``Conservative'' mode verifies the checksum for
every object access (including those read by \texttt{pgl\_get} without
\mbufing{}).

\reftab{tab:ftmodel} quantifies the vulnerability using the amount of object
data that is accessed without checksum verification. The data accumulates
across all transactions for \Po{}, \FTLib{}-MLPC, and
``Conservative'' modes. For ``Scrub'' modes, we count the vulnerable data
between two scrubbing runs. Numbers in \reftab{tab:ftmodel} are normalized to
\Po{}, which does not have any checksum protection for object data.

The cost of verifying checksums for every object access depends strongly on the
data structure size and its insertion algorithm. For small objects, such as
ctree, rbtree, and hashmap, the cost is negligible. But for btree, skiplist,
and rtree, due to their large object sizes, the cost is significant. Thus, a
scrubbing-based policy could be faster, with more data subject to
corruption between two successive runs.

\subsection{Error Detection and Correction}

\FTLib{} provides error injection functions to emulate both
hardware-uncorrectable NVMM media errors and hardware-undetectable scribbles.

We initially developed \FTLib{} using conventional DRAM machines that lack
support for injecting NVMM errors at the hardware level. Therefore, we use
\texttt{mprotect()} and \texttt{SIGSEGV} to emulate NVMM media errors and
\texttt{SIGBUS}. When an NVMM file is \DAXmmapd{}, the injector can randomly choose a
page that contains user-allocated objects, erase it, and call
\texttt{mprotect(PROT\_NONE)} on the page. Later, when the application reads the
corrupted page, \FTLib{} intercepts \texttt{SIGSEGV}, changes the page to
read/write mode, and restores the page's data.
The injector function can also randomly corrupt a metadata region or a victim
object to emulate software-induced, scribble errors.

In both test cases, we observe \FTLib{} can successfully repair a victim page or
an object and resume normal program execution.
In our evaluation using a 100~GB pool and 1~GB parity, we measured 180 $\mu$s to
repair a page of a page column.

We also intentionally introduce buffer overrun bugs in our applications and
observe that \FTLib{} can successfully detect them using \mbuf{} canaries.
The transaction then aborts to prevent any NVMM corruption. We have also
verified \FTLib{} is compatible with AddressSanitizer for detecting buffer
overrun bugs (when updating a \mbufed{} object exceeds its buffer's boundary),
if both \FTLib{} and its application code are compiled with support.

\section{Related Work} \label{sec:related}

In this section, we place \FTLib{} in context relative to  previous projects
that have explored how to use NVMM effectively.

\boldparagraph{Transaction Support} All previous libraries for using NVMMs to
build complex objects rely on transactions for crash consistency. Although we
built \FTLib{} on \lpo{}, its techniques could be applied to another persistent
object system. NV-Heaps~\cite{nvheaps}, Atlas~\cite{atlas}, DCT~\cite{dct},
and \lpo{}~\cite{pmdk} provide undo logging for applications to snapshot
persistent objects before making in-place updates. Mnemosyne~\cite{mnemosyne},
SoftWrAp~\cite{softwrap}, and DUDETM~\cite{dudetm} use variations of redo
logging. REWIND~\cite{rewind} implements both undo and redo logging for
fine-grained, high-concurrent transactions. Log-structured NVMM~\cite{lognvmm}
makes changes to objects via append-only logs, and it does not require extra
logging for consistency. Romulus~\cite{romulus} uses a main-back mechanism to
implement efficient redo log-based transactions.

None of these systems provide fault tolerance for NVMM errors. We believe they
can adopt \FTLib{}'s parity and checksum design to improve their resilience to
NVMM errors at low storage overhead. In \refsec{sec:update} we
described how to apply the hybrid parity updating scheme to an undo
logging-based system. Log-structured and copy-on-write systems can adopt the
techniques in similar ways.

\boldparagraph{Fault Tolerance} Both \FTLib{} and \lpo{}'s replication mode
protect against media errors, but \FTLib{} provides stronger
protection and much lower space overhead. Furthermore, \lpo{} can only repair
media errors offline, and it does not detect or repair software corruption
to user objects.

NVMalloc~\cite{nvmalloc} uses checksums to protect metadata. It does not specify
whether application data is also checksum-protected, and it does not provide any
form of redundancy to repair the corruption. NVMalloc uses \texttt{mprotect()} to
protect NVMM pages while they are not mapped for writing. \FTLib{} could adopt
this technique to prevent an application from scribbling its own persistent data
structures.

The NOVA file system~\cite{nova, novafortis} uses parity-based protection for
file data.  However, it must disable these features for NVMM pages that are
\DAXmmapd{} for writing in user-space, since the page's contents can change
without the file system's knowledge, making it impossible for NOVA to keep the
parity information consistent if an application modifies \DAXmmapd{} data.
As a result, Pangolin's and NOVA's fault tolerance mechanisms are
complementary.

\section{Conclusion} \label{sec:conclusion}

This work presents \FTLib{}, a fault-tolerant, \DAXmmapd{} NVMM programming
library for applications to build complex data structures in NVMM.
\FTLib{} uses a novel, space-efficient layout of data
and parity to protect arbitrary-sized NVMM objects combined with per-object
checksums to detect corruption. To maintain high performance, \FTLib{} uses
\mbufing{}, carefully-chosen parity and checksum updating algorithms.
As a result, \FTLib{} provides stronger protection,
better availability, and much
lower storage overhead than existing NVMM programming libraries.

{\normalsize \bibliographystyle{plain}
\bibliography{paper}}
\end{document}